\documentclass{emulateapj}
\usepackage{apjfonts}
\usepackage{rotating}
\usepackage{amsmath}
\newcommand{\pivec}{\mbox{\boldmath $\pi$}}
\newcommand{\muvec}{\mbox{\boldmath $\mu$}}

\lefthead{SKOWRON ET AL.} 
\righthead{MICROLENSING PLANET AROUND M DWARF}

\begin{document}
\title{OGLE-2011-BLG-0265L\MakeLowercase{b}: a Jovian Microlensing  
Planet Orbiting an M Dwarf}

\author{
J.~Skowron$^{O1,O}$, I.-G.~Shin$^{U1}$, A.~Udalski$^{O1,O}$, C.~Han$^{U1,U,\star}$, T.~Sumi$^{M1,M}$, 
Y.~Shvartzvald$^{W1,W}$, A.~Gould$^{U2,U}$, D.~Dominis~Prester$^{P7,P}$, R.~A.~Street$^{R1,R}$, 
U.~G.~J{\o}rgensen$^{S1,S2,S}$, D.~P.~Bennett$^{M3,M}$, V.~Bozza$^{S4,S5,S}$\\
and\\
% OGLE -----------------------------
M.~K.~Szyma{\'n}ski$^{O1}$, M.~Kubiak$^{O1}$, G.~Pietrzy{\'n}ski$^{O1,O2}$, 
I.~Soszy{\'n}ski$^{O1}$, R.~Poleski$^{O1,U2}$,
S.~Koz{\l}owski$^{O1}$, P.~Pietrukowicz$^{O1}$,
K.~Ulaczyk$^{O1}$, {\L}.~Wyrzykowski$^{O3,O1}$\\
(The OGLE Collaboration),\\
% MOA ------------------------------
F.~Abe$^{M2}$, A.~Bhattacharya$^{M3}$, I.~A.~Bond$^{M4}$, C.~S.~Botzler$^{M5}$, 
M.~Freeman$^{M5}$, A.~Fukui$^{M7}$, D.~Fukunaga$^{M2}$, Y.~Itow$^{M2}$, 
C.~H.~Ling$^{M4}$, N.~Koshimoto$^{M1}$, K.~Masuda$^{M2}$, Y.~Matsubara$^{M2}$, 
Y.~Muraki$^{M8}$, S.~Namba$^{M1}$, K.~Ohnishi$^{M9}$, L.~C.~Philpott$^{M11}$, 
N.~Rattenbury$^{M5}$, T.~Saito$^{M10}$, D.~J.~Sullivan$^{M6}$, D.~Suzuki$^{M1}$, 
P.~J.~Tristram$^{M6}$, P.~C.~M.~Yock$^{M5}$\\
(The MOA Collaboration),\\
% Wise ------------------------------
D.~Maoz$^{W1}$, S.~Kaspi$^{W1}$, M.~Friedmann$^{W1}$\\
(Wise group),\\
% uFUN ------------------------------
L.~A.~Almeida$^{U8}$, V.~Batista$^{U2}$, G.~Christie$^{U3}$, J.-Y.~Choi$^{U1}$, 
D.~L.~DePoy$^{U4}$, B.~S.~Gaudi$^{U2}$, C.~Henderson$^{U2}$, K.-H.~Hwang$^{U1}$,
F.~Jablonski$^{U8}$, Y.~K.~Jung$^{U1}$, C.-U.~Lee$^{U5}$, J.~McCormick$^{U6}$, 
T.~Natusch$^{U3,U7}$, H.~Ngan$^{U3}$, H.~Park$^{U1}$, R.~W.~Pogge$^{U2}$, J.~C.~Yee$^{U2,U9}$\\
(The $\mu$FUN Collaboration)\\
% PLANET ------------------------------
M.~D.~Albrow$^{P2}$, E.~Bachelet$^{P3}$, J.-P.~Beaulieu$^{P1}$, S.~Brillant$^{P4}$, 
J.~A.~R.~Caldwell$^{P5}$, A.~Cassan$^{P1}$, A.~Cole$^{P6}$, E.~Corrales$^{P1}$, 
Ch.~Coutures$^{P1}$, S.~Dieters$^{P3}$, J.~Donatowicz$^{P11}$, P.~Fouqu\'e$^{P3,P8}$, 
\fbox{J. Greenhill}$^{P6,\dagger}$, N.~Kains$^{P9,S17,R,S}$,
S.~R.~Kane$^{P10}$, D.~Kubas$^{P1,P4}$, J.-B.~Marquette$^{P1}$, 
R.~Martin$^{P12}$, J.~Menzies$^{P13}$, K.~R.~Pollard$^{P2}$, C.~Ranc$^{P1}$, 
K.~C.~Sahu$^{S17,S}$, J.~Wambsganss$^{P14,S}$, A.~Williams$^{P14}$, D.~Wouters$^{P1}$\\
(The PLANET Collaboration)\\
% RoboNet --------------------------
Y.~Tsapras$^{R1,R5,P}$, D.~M.~Bramich$^{R2}$, K.~Horne$^{R3,P}$, M.~Hundertmark$^{R3,S9,S}$, 
%S. Ipatov$^{R4}$, 
C.~Snodgrass$^{R4,S}$, I.~A.~Steele$^{R6}$\\
(The RoboNet Collaboration)\\
% MiNDSTEp -------------------------$cv   
K.~A.~Alsubai$^{S3}$, P.~Browne$^{R3,R}$, M.~J.~Burgdorf$^{S6}$, 
S.~Calchi~Novati$^{P10,S4,S8,\ast}$, P.~Dodds$^{R3}$, M.~Dominik$^{R3,R}$, 
S.~Dreizler$^{S9}$, X.-S.~Fang$^{S22}$, C.-H.~Gu$^{S22}$, 
Hardis$^{S1}$, K.~Harps{\o}e$^{S1,S2}$, 
F.~V.~Hessman$^{S9}$, T.~C.~Hinse$^{U5,S1,S11}$, A.~Hornstrup$^{S12}$,
J.~Jessen-Hansen$^{S10}$, E.~Kerins$^{S12}$, C.~Liebig$^{R3}$, 
M.~Lund$^{S10}$, M.~Lundkvist$^{S10,S21}$, L.~Mancini$^{S13}$, 
M.~Mathiasen$^{S1}$, 
M.~T.~Penny$^{S12,U2}$, S.~Rahvar$^{S14,S15}$, D.~Ricci$^{S16,S19,S20}$, 
G.~Scarpetta$^{S4,S5,S8}$, J.~Skottfelt$^{S1,S2}$,
J.~Southworth$^{S18}$, J.~Surdej$^{S16}$, J.~Tregloan-Reed$^{S7}$, 
O.~Wertz$^{S16}$\\
(The MiNDSTEp consortium)\\
}
% -----------------------------------------------------------------------------------------------------------
\affil{$^{O1}$ Warsaw University Observatory, Al. Ujazdowskie 4, 00-478 Warszawa, Poland}
\affil{$^{O2}$ Universidad de Concepci\'on, Departamento de Astronomia, Casilla 160-C, Concepci\'on, Chile}
\affil{$^{O3}$ Institute of Astronomy, University of Cambridge, Madingley Road, Cambridge CB3 0HA, UK}
% ===========================================================
\affil{$^{U1}$ Department of Physics, Institute for Astrophysics, Chungbuk National University, 371-763 Cheongju, Korea}
\affil{$^{U2}$ Department of Astronomy, Ohio State University, 140 West 18th Avenue, Columbus, OH 43210, USA}
\affil{$^{U3}$ Auckland Observatory, 670 Manukau Rd, Royal Oak 1023, Auckland, New Zealand}
\affil{$^{U4}$ Dept. of Physics and Astronomy, Texas A\&M University College Station, TX 77843-4242, USA}
\affil{$^{U5}$ Korea Astronomy and Space Science Institute, 305-348 Daejeon, Korea}
\affil{$^{U6}$ Farm Cove Observatory, Centre for Backyard Astrophysics, Pakuranga, Auckland, New Zealand}
\affil{$^{U7}$ Institute for Radiophysics and Space Research, AUT University, Auckland, New Zealand}
\affil{$^{U8}$ Divis\~{a}o de Astrofisica, Instituto Nacional de Pesquisas Espeaciais, Avenida dos Astronauntas, 1758 Sao Jos\'e dos Campos, 12227-010 SP, Brazil}
\affil{$^{U9}$ Harvard-Smithsonian Center for Astrophysics, 60 Garden St., Cambridge, MA 02138, USA}
% ===========================================================  
\affil{$^{M1}$ Dept. of Earth and Space Science, Graduate School of Science, Osaka University, 1-1 Machikaneyama-cho, Toyonaka, 560-0043 Osaka, Japan} 
\affil{$^{M2}$ Solar-Terrestrial Environment Laboratory, Nagoya University, 464-8601 Nagoya, Japan} 
\affil{$^{M3}$ Dept. of Physics, University of Notre Dame, Notre Dame, IN 46556, USA} 
\affil{$^{M4}$ Institute of Information and Mathematical Sciences, Massey University, Private Bag 102-904, North Shore Mail Centre, Auckland, New Zealand} 
\affil{$^{M5}$ Dept. of Physics, University of Auckland, Private Bag 92019, Auckland, New Zealand} 
\affil{$^{M6}$ School of Chemical and Physical Sciences, Victoria University, Wellington, New Zealand} 
\affil{$^{M7}$ Okayama Astrophysical Observatory, National Astronomical Observatory of Japan, Asakuchi, 719-0232 Okayama, Japan} 
\affil{$^{M8}$ Dept. of Physics, Konan University, Nishiokamoto 8-9-1, 658-8501 Kobe, Japan} 
\affil{$^{M9}$ Nagano National College of Technology, 381-8550 Nagano, Japan} 
\affil{$^{M10}$ Tokyo Metropolitan College of Industrial Technology, 116-8523 Tokyo, Japan} 
\affil{$^{M11}$ Department of Earth, Ocean and Atmospheric Sciences, University of British Columbia, Vancouver, British Columbia, V6T 1Z4, Canada}
% ===========================================================
\affil{$^{P1}$ UPMC-CNRS, UMR7095, Institut d'Astrophysique de Paris, 98bis boulevard Arago, 75014 Paris, France} 
\affil{$^{P2}$ University of Canterbury, Dept. of Physics and Astronomy, Private Bag 4800, 8020 Christchurch, New Zealand} 
\affil{$^{P3}$ Universit\'e de Toulouse, UPS-OMP, IRAP, 31400 Toulouse, France} 
\affil{$^{P4}$ European Southern Observatory (ESO), Alonso de Cordova 3107, Casilla 19001, Santiago 19, Chile} 
\affil{$^{P5}$ McDonald Observatory, 16120 St Hwy Spur 78 \#2, Fort Davis, TX 79734, USA} 
\affil{$^{P6}$ School of Math and Physics, University of Tasmania, Private Bag 37, GPO Hobart, 7001 Tasmania, Australia} 
\affil{$^{P7}$ Physics Department, Faculty of Arts and Sciences, University of Rijeka, Omladinska 14, 51000 Rijeka, Croatia} 
\affil{$^{P8}$ CFHT Corporation, 65-1238 Mamalahoa Hwy, Kamuela, HI, 96743, USA} 
\affil{$^{P9}$ European Southern Observatory, Karl-Schwarzschild-Str. 2, 85748 Garching bei M\"{u}nchen, Germany} 
\affil{$^{P10}$ NASA Exoplanet Science Institute, Caltech, MS 100-22, 770 S.  Wilson Ave., Pasadena, CA 91125, USA} 
\affil{$^{P11}$ Technical University of Vienna, Department of Computing,Wiedner Hauptstrasse 10, Vienna, Austria} 
\affil{$^{P12}$ Perth Observatory, Walnut Road, Bickley, 6076 Perth, Australia} 
\affil{$^{P13}$ South African Astronomical Observatory, PO Box 9, Observatory 7935, South Africa} 
\affil{$^{P14}$ Astronomisches Rechen-Institut, Zentrum f\"{u}r Astronomie der Universit\"{a}t Heidelberg (ZAH), M\"{o}nchhofstr. 12-14, 69120 Heidelberg, Germany}
% ===========================================================
\affil{$^{W1}$ School of Physics and Astronomy, Tel-Aviv University, Tel-Aviv 69978, Israel} 
% ===========================================================
\affil{$^{R1}$ Las Cumbres Observatory Global Telescope Network, 6740 Cortona Drive, suite 102, Goleta, CA 93117, USA}
\affil{$^{R2}$ Qatar Environment and Energy Research Institute, Qatar Foundation, P. O. Box 5825, Doha, Qatar} 
\affil{$^{R3}$ SUPA School of Physics \& Astronomy, University of St Andrews, North Haugh, St Andrews, KY16 9SS, UK} 
\affil{$^{R4}$ Max Planck Institute for Solar System Research, Justus-von-Liebig-Weg 3, 37077 G\"{o}ttingen, Germany}
\affil{$^{R5}$ School of Physics and Astronomy, Queen Mary University of London, Mile End Road, London E1 4NS, UK}
\affil{$^{R6}$ Astrophysics Research Institute, Liverpool John Moores University, Liverpool CH41 1LD, UK}
% ===========================================================
\affil{$^{S1}$ Niels Bohr Institute, University of Copenhagen, Juliane Maries vej 30, 2100 Copenhagen, Denmark} 
\affil{$^{S2}$ Centre for Star and Planet Formation, Geological Museum, {\O}ster Voldgade 5, 1350 Copenhagen, Denmark} 
\affil{$^{S3}$ Qatar Foundation, PO Box 5825, Doha, Qatar} 
\affil{$^{S4}$ Dipartimento di Fisica "E.R. Caianiello" Universit$\grave{\rm a}$ degli Studi di Salerno Via Giovanni Paolo II - I 84084 Fisciano (SA) - Italy}
\affil{$^{S5}$ Istituto Nazionale di Fisica Nucleare, Sezione di Napoli, Italy} 
\affil{$^{S6}$ HE Space Operations GmbH, Flughafenallee 24, 28199 Bremen, Germany}
\affil{$^{S7}$ NASA Ames Research Center, Moffett Field CA 94035, USA} 
\affil{$^{S8}$ Istituto Internazionale per gli Alti Studi Scientifici (IIASS), Via Giuseppe Pellegrino, 19, 84019 Vietri Sul Mare Salerno, Italy} 
\affil{$^{S9}$ Institut f\"{u}r Astrophysik, Georg-August-Universit{\"{a}}t, Friedrich-Hund-Platz 1, 37077 G\"{o}ttingen, Germany} 
\affil{$^{S10}$ Department of Physics and Astronomy, Aarhus University, Ny Munkegade 120, $\AA$rhus C, Denmark} 
\affil{$^{S11}$ Armagh Observatory, College Hill, Armagh, BT61 9DG, Northern Ireland, UK} 
\affil{$^{S12}$ Jodrell Bank Centre for Astrophysics, University of Manchester, Oxford Road, Manchester, M13 9PL, UK} 
\affil{$^{S13}$ Max Planck Institute for Astronomy, K\"{o}nigstuhl 17, 69117 Heidelberg, Germany} 
\affil{$^{S14}$ Department of Physics, Sharif University of Technology, PO Box 11155--9161, Tehran, Iran} 
\affil{$^{S15}$ Perimeter Institute for Theoretical Physics, 31 Caroline St. N., Waterloo ON, N2L 2Y5, Canada} 
\affil{$^{S16}$ Institut d'Astrophysique et de G\'eophysique, All\'e du 6 Ao$\hat{\rm u}$t 17, Sart Tilman, B$\hat{\rm a}$t. B5c, 4000 Li\'ege, Belgium} 
\affil{$^{S17}$ Space Telescope Science Institute, 3700 San Martin Drive, Baltimore, MD 21218, USA} 
\affil{$^{S18}$ Astrophysics Group, Keele University, Staffordshire, ST5 5BG, UK } 
\affil{$^{S19}$ INAF/Istituto di Astrofisica Spaziale e Fisica Cosmica, Via Gobetti 101, Bologna, Italy}
\affil{$^{S20}$ Instituto de Astronom\'ia, UNAM, AP 877, Ensenada, B.C. 22800, Mexico}
\affil{$^{S21}$ Stellar Astrophysics Centre, Department of Physics and Astronomy, Aarhus University, Ny Munkegade 120, DK-8000 Aarhus C, Denmark}
%\affil{$^{S22}$ NASA Exoplanet Science Institute, MS 100-22, California Institute of  Technology, Pasadena, CA 91125, USA}
% ------------------------------------------------------------------------------------------------------------
\affil{$^{\star}$ Corresponding author}
\affil{$^{\dagger}$ deceased 28 September 2014}
\affil{$^{\ast}$ Sagan Visiting Fellow}
\affil{$^{O}$ The OGLE Collaboration}
\affil{$^{M}$ The MOA Collaboration}
\affil{$^{W}$ The Wise Group}
\affil{$^{U}$ The $\mu$FUN Collaboration}
\affil{$^{P}$ The PLANET Collaboration}
\affil{$^{R}$ The RoboNet Collaboration}
\affil{$^{S}$ The MiNDSTEp Consortium}

\begin{abstract}

We report the discovery of a Jupiter-mass planet orbiting an M-dwarf star 
that gave rise to the microlensing event OGLE-2011-BLG-0265.  Such a system 
is very rare among known planetary systems and thus the discovery is 
important for theoretical studies of planetary formation and evolution.
High-cadence temporal coverage of the planetary signal combined with 
extended observations throughout the event allows us to accurately model 
the observed light curve. The final microlensing solution remains, however,
degenerate yielding two possible configurations of the planet and the
host star.  In the case of the preferred solution, the mass of the planet is 
$M_{\rm p} = 0.9\pm 0.3\ M_{\rm J}$, and the planet is orbiting a star with 
a mass $M = 0.22\pm 0.06\ M_\odot$. The second possible configuration 
(2$\sigma$ away) consists of a planet with $M_{\rm p}=0.6\pm 0.3\ M_{\rm J}$ 
and host star with $M=0.14\pm 0.06\ M_\odot$. The system is located in the 
Galactic disk 3 -- 4 kpc towards the Galactic bulge. In both cases, with 
an orbit size of 1.5 -- 2.0 AU, the planet is a ``cold Jupiter'' -- located 
well beyond the ``snow line'' of the host star.  Currently available data 
make the secure selection of the correct solution difficult, but there are 
prospects for lifting the degeneracy with additional follow-up observations 
in the future, when the lens and source star separate.

\end{abstract}


\section{Introduction}

In the recent decade, gravitational lensing has proven to be one of the 
major techniques of detecting and characterizing extrasolar planetary 
systems. Due to the favorable geometry in the Galaxy where microlensing 
phenomena occur, this technique is sensitive to planets orbiting their 
host stars with separations 0.5--10 AU. The technique is sensitive to 
low-mass planets -- down to Earth-mass planets and even smaller masses 
if observed from space. It can also detect planets not bound to stars 
-- free-floating planets \citep{sumi11}. Therefore, it provides an 
important tool that enables a census of extrasolar  planets in the very 
important region of parameter space that is generally inaccessible to 
other techniques: the region beyond the snow line where cold giant 
planets are most probably forming. Such a census will be complementary 
to the one provided by transit and radial-velocity surveys.

First assessments of the planet frequency in the microlensing domain
have already been published \citep{tsapras03,gould10, sumi10, cassan12}.
However, these studies were based on a limited number of planetary 
microlensing events. Precise analysis requires a much larger number of 
microlensing planets. New observational strategies of microlensing 
experiments have been implemented in the last several years, leading 
to significant increase of the number of planet detections.

After the initial period of pioneering detections, the planetary 
microlensing field has undergone rapid changes and continues to evolve 
toward the next-generation experiments. The traditional first-generation
approach was that some selected microlensing events detected by
large-scale surveys like the Optical Gravitational Lensing Experiment
(OGLE) and the Microlensing Observation in Astrophysics (MOA) projects
were densely observed by follow-up groups such as $\mu$FUN, PLANET,  
RoboNet and MiNDSTEp. Since then, the experiments have adopted more 
sophisticated observing strategies. For example, the second-generation 
microlensing surveys consist of a network of wide-field telescopes 
capable of observing large areas of the Galactic bulge field with 
high cadences of about 10 -- 20 minutes. Starting from the 2010 observing 
season when the fourth phase of the OGLE survey began regular observations 
with the 1.3-m telescope at the Las Campanas Observatory in Chile, the 
second-generation microlensing network began to take shape. The OGLE-IV 
observing setup together with the 1.8-m MOA-II telescope located at Mount 
John Observatory in New Zealand and the 1-m telescope at the Wise 
Observatory in Israel became the backbone of the second-generation network 
capable of conducting round-the-clock observations of selected fields
in the Galactic bulge. There has also been progress in follow-up observations, 
including the formation of new-generation follow-up networks with enhanced 
observing capability, e.g., RoboNet (a network of robotic telescopes 
from LCOGT and the Liverpool Telescope).

One of the most important discoveries made with the microlensing technique
is the detection of cold giant planets orbiting faint M-type dwarf stars. 
These discoveries are the straight consequence of the fact that microlensing 
does not rely on the light from a host star in order to detect a planet. 
This implies that the dependency of the microlensing sensitivity to planets 
on the spectral type of host stars is weak and the sensitivity extends down 
to late M~dwarfs and beyond.

Studying planets around M~dwarfs is important because these stars comprise 
$\sim 70\%$  -- 75\% of stars in the Solar neighborhood and the Galaxy as 
a whole.  Planets around M~dwarfs have been probed by the radial-velocity 
and  transit methods,  e.g., \citet{delfosse98}, \citet{marcy98}, 
\citet{bonfils11}, \citet{montet14},  and \citet{charbonneau09}.  However, 
the low luminosity of M~dwarfs poses serious difficulties in searching for 
planets with these methods.  Furthermore, the host stars of M-dwarf planets 
discovered so far tend to occupy the brighter end of the M~dwarf range. 
As a result, the characteristics of the lower-mass M-dwarf planet population 
are essentially unknown. In addition, all M-dwarf planets detected by the 
radial-velocity method are located within only a few dozens of parsecs from 
the Sun and thus the sample of these planets is greatly biased not only to 
the spectral type of host stars but also to the distance from the Solar system.

By contrast, the most frequent host stars of microlensing planets are M
dwarfs, including a planet with its host star directly imaged 
\citep{bennett08, kubas12} and several others whose masses are constrained  
by microlensing light curves and auxiliary data \citep{udalski05, dong09, 
beaulieu06, gaudi08, bennett10, batista11, street13, kains13, poleski14, 
tsapras14, shvartzvald14}.  In addition, lensing events occur regardless 
of the stellar types of lensing objects and thus one can obtain a sample 
of planetary systems unbiased by the stellar types of host stars.  
Furthermore, lensing events occur by objects distributed in a wide range 
of the Galaxy between the Earth and the Galactic center and thus one can
obtain a planet sample more representative of the whole Galaxy.

Constructing an unbiased sample of planets around M~dwarfs is important 
for understanding the formation mechanism of these planets.  A theory 
based  on the core accretion mechanism predicts that gas giants form much 
less frequently around M~dwarfs than around Sun-like stars, while 
terrestrial and ice giant planets may be relatively common 
\citep{laughlin04, ida05}.  An alternative theory based on the disk 
instability mechanism predicts that  giant planets can form around M
dwarfs \citep{boss06} -- the opposite to the prediction of planet
formation by the core accretion mechanism. Therefore, determining the
characteristics and the frequency of planets orbiting M~dwarfs is important 
in order to refine the planetary formation scenario  of these planets.

In this paper, we report the discovery of another giant planet orbiting 
an M3-M4 dwarf that was detected from the light curve analysis of the 
microlensing event OGLE-2011-BLG-0265. Although modeling the 
microlensing light curve yields two solutions that cannot be fully
distinguished with the currently available data, both solutions 
indicate a Jupiter-mass planet.  There is good prospect on resolving the 
ambiguity of the solutions in the future when the lens and the source 
separate.

\section{OBSERVATIONS AND DATA}

The gravitational microlensing event OGLE-2011-BLG-0265 was discovered
on 2011 April 16 by the OGLE Early Warning System (EWS) during the
test phase of its implementation for the OGLE-IV survey. It was
officially  announced on 2011 May 25 as one of 431 events in the
inauguration set of  events detected during the 2011 season. The event
was also found by the MOA group and designated as MOA-2011-BLG-197.

\begin{figure*}[ht]
%\epsscale{0.99}
\epsscale{0.80}
\plotone{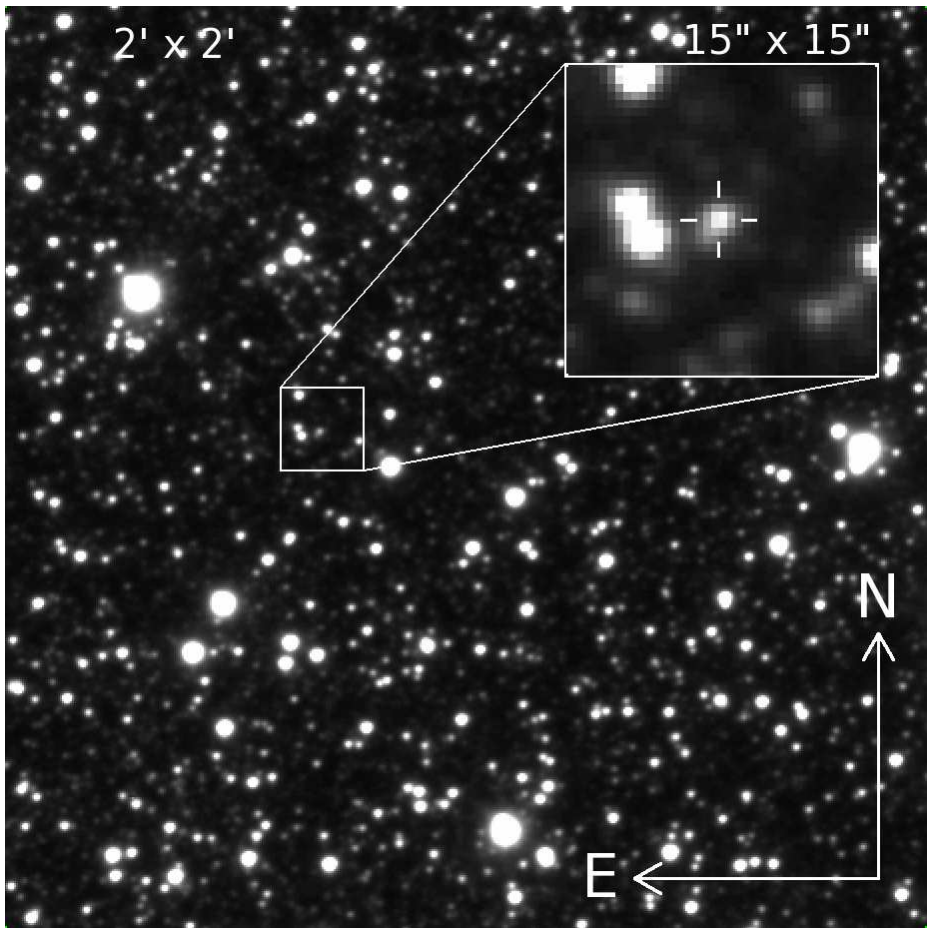}
\caption{\label{fig:one}
Finding chart for the microlensing event OGLE-2011-BLG-0265 as seen 
at the baseline level in 2010.  The position of the source star (and 
the lens) is marked with the white cross, $(\alpha,\delta)_{\rm J2000}$ 
= $(17^{\rm h}57^{\rm m}47.72^{\rm s}$, $-27^\circ 23' 40''\hskip-2pt .3)$ 
$\pm 0.1''$.  The field of view is 2' x 2', while the inset covers 
15" x 15". Pixel scale is 0.26''/px.  North is up and east is to the left. 
The brightest stars in the inset are $I \sim 16.2-16.4$, while the faintest 
visible at this scale are $I \sim 20$.  The brightest star in the whole 
chart is TYC 6849-852-1 ($I \approx 11$).}
\end{figure*}

The microlensed source star of the event is located at
$(\alpha,\delta)_{\rm J2000} =(17^{\rm h}57^{\rm m}47.72^{\rm s},
-27^\circ 23' 40''\hskip-2pt .3)$  in equatorial coordinates and
$(l,b)=(2.70^\circ,-1.52^\circ)$ in Galactic  coordinates (with the 
accuracy of the absolute position of the order of 0.1 arcsec). This 
region of the sky corresponds to the densest stellar region in the 
Galactic bulge toward which vast majority of microlensing events are 
being detected.  Figure~\ref{fig:one} shows the finding chart of the 
event taken in 2010 when the source had not yet been magnified.  The 
brightness and color of the event at the baseline, calibrated to the 
standard $VI$ system, are $I=17.51$ and $V-I=3.03$, respectively.

% Figure 1 ------------------------------------------------------
% --------------------------------------------------------------

The OGLE-IV survey is conducted using the 1.3-m Warsaw telescope
equipped with the 32-CCD mosaic camera located at the Las Campanas
Observatory in Chile. A single image covers approximately 1.4 square
degrees with a resolution of 0.26 arcsec/pixel. OGLE-2011-BLG-0265 is
located in the ``BLG504'' OGLE-IV field which was observed with 
18-minute cadence in the 2011 season. See  the OGLE Web page\footnote{\tt
http://ogle.astrouw.edu.pl/sky/ogle4-BLG/} for the map of  the sky
coverage. The exposure time was 100 seconds and the variability
monitoring  was performed in the $I$-band filter. Several $V$-band
images were also taken during the event in order to determine the color
of the source star. The analyzed OGLE-IV data set of the event contains
3749 epochs covering three observing  seasons 2010 -- 2012.

The MOA project is regularly surveying the Galactic bulge with the 1.8-m
telescope at the Mt. John Observatory in New Zealand. Images are
collected with a ten-CCD mosaic camera covering $\approx 2.2$ square
degrees. OGLE-2011-BLG-0265 lies in the high-cadence MOA field ``gb10''
which is typically visited a few times per hour, enabling to take 4774
epochs in total  during the 2006-2012 seasons. Observations were
conducted using the wide non-standard $R$/$I$ filter with the exposure 
time of 60 seconds.

OGLE-2011-BLG-0265 is also located in the footprint of the survey
conducted at the Wise Observatory in Israel with the 1.0-m telescope
and four-CCD mosaic camera, LAIWO \citep{shvartzvald12}. This site 
fills the longitudinal gap between the OGLE and MOA sites enabling 
round-the-clock coverage of the event. In total 710 epochs were 
obtained from this survey.  Observations were carried out with the 
$I$-band filter and the exposure time was 180 seconds.

OGLE-2011-BLG-0265 event turned out to evolve relatively slowly.
Data collected by survey observations have good enough coverage of 
the anomaly and overall light curve to identify the planetary nature 
of the event. Nevertheless, the phenomenon was also monitored by 
several follow-up groups based on the anomaly alert issued on July 2, 
2011 by the MOA group. It should be noted that the OGLE group generally 
does not issue alerts of ongoing anomalies in the present phase.

The groups that participated in the follow-up observations include 
the Probing Lensing Anomalies NETwork (PLANET: Beaulieu et al. 2006), 
Microlensing Follow-Up Network ($\mu$FUN: Gould et al. 2006), RoboNet 
\citep{tsapras09}, and MiNDSTEp \citep{dominik10}.  Telescopes used 
for these observations include PLANET 1.0-m of South African Astronomical 
Observatory (SAAO) in South Africa, PLANET 0.6-m of Perth Observatory 
in Australia, $\mu$FUN 1.3-m SMARTS telescope of Cerro Tololo 
Inter-American Observatory (CTIO) in Chile, $\mu$FUN 0.4-m of Auckland 
Observatory in New Zealand, $\mu$FUN 0.36-m of Farm Cove Observatory 
(FCO) in New Zealand, $\mu$FUN 0.8-m of Observatorio del Teide in 
Tenerife, Spain, $\mu$FUN 0.6-m of Observatorio do Pico dos Dias (OPD) 
in Brazil, $\mu$FUN 0.4-m of Marty S. Kraar Observatory of Weizmann 
Institute of Science (Weizmann) in Israel, MiNDSTEp Danish 1.54-m 
telescope at La Silla Observatory in Chile, 2.0-m Liverpool Telescope 
at La Palma, RoboNet FTN 2.0-m in Hawaii, and RoboNet FTS 2.0-m in 
Australia.

By the time the first anomaly had ended, a series of solutions of lensing
parameters  based on independent real-time modeling were released. A
consistent interpretation of these analyses was that the anomaly was
produced by a planetary companion to the lens star. The models also
predicted that there would be another perturbation in about ten days
after the first anomaly followed by the event peak just after the
second anomaly. Based on this prediction, follow-up observations were
continued beyond the main anomaly up to the peak  and even beyond. This
enabled dense coverage of the second anomaly which turned out  to be
important for the precise characterization of the lens system. 
See Section \ref{sec:solution}.

The event did not return to its baseline until the end of the 2011
season --  ${\rm HJD}'(={\rm HJD}-2450000)\sim 5870$. In order to obtain
baseline data, observations were resumed in the 2012 season that
started  on ${\rm HJD}'\sim 5960$. Combined survey and follow-up
photometry constitute  a very continuous and complete data set with the
very dense coverage of the planetary anomaly.

% Figure 2 -----------------------------------------------------------------------------------------
\begin{figure*}[ht]
%\epsscale{0.99}
\epsscale{0.80}
\plotone{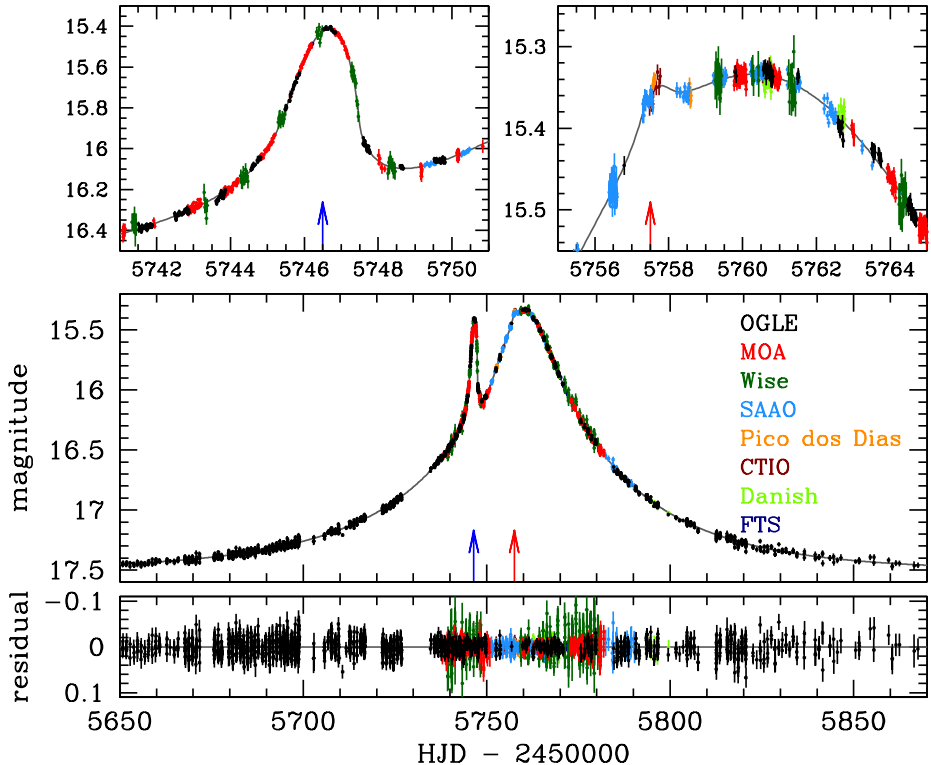}
\caption{\label{fig:two}
Light curve of the microlensing event OGLE-2011-BLG-0265.  Only a subset 
of data taken during the event is presented -- lower signal-to-noise 
observations, as well as, data providing little constraint on the final 
solution are omitted.  Colors of data points are chosen to match those 
of the labels of  observatories.  The solid curve superposed on the data 
points represents the best-fit model curve ($u_0>0$). The model curve for 
the $u_0<0$ solution would be visually indistinguishable. The two upper
panels show the enlarged view of the major (${\rm HJD} \sim 2455746.5$)
and minor ($\sim 2455757.5$) planetary perturbation regions (marked with
arrows).  }
\end{figure*}
% ---------------------------------------------------------------------------------------------------

Data acquired from different observatories were reduced using photometry
codes that were developed by the individual groups. The photometry codes 
used by the OGLE and MOA groups, developed respectively by \citet{udalski03} 
and \citet{bond01}, are based on the Difference Image Analysis method 
\citep{alard98}. The PySIS pipeline \citep{albrow09} was used for the 
reduction of the PLANET data and the Wise data. The $\mu$FUN data were 
processed using the DoPHOT pipeline \citep{schechter93}.  For the RoboNet 
and MiNDSTEp data, the DanDIA pipeline \citep{bramich08} was used.

To analyze the data sets obtained from different observatories, we
rescale the reported uncertainties for each data set \cite[cf.][]{skowron11}. 
The microlensing magnification significantly changes the brightness of the 
measured object during the event and it is often the case that the reported 
uncertainties by the automatic pipelines are underestimated by different 
amounts.  To account for this, we first adjust uncertainties by introducing 
a quadratic term so that the cumulative distribution function of $\chi^2$ 
as a function of magnification becomes linear. We then rescale error bars 
so that $\chi^2$ per degree of freedom (dof) becomes unity for each data 
set, where the value of $\chi^2$ is derived from the best-fit solution. 
This process greatly helps to estimate uncertainties of the lensing parameters. 
It is done in an iterative manner using the full model (i.e., with effects of 
parallax and orbital motion taken into account).

Figure \ref{fig:two} shows the light curve of OGLE-2011-BLG-0265.  The 
subset of gathered data that were used in the final calculations is presented.  
For the most part, the light curve is well represented by a smooth and
symmetric curve of a standard lensing event caused by a single-mass object 
\citep{paczynski86} except for the short-term perturbations at 
${\rm HJD}'\sim 5746.5$ (major  perturbation) and 5757.5 (minor
perturbation), which lasted for $\sim 4$ days  and $\sim 1$ day,
respectively. These short-term perturbations are characteristic 
features of planetary microlensing \citep{mao91, gould92b}.

The dense temporal coverage from the multiple sites is useful in ensuring 
that there is no missing feature in the light curve.  Also, overlapping 
observations allows to perform extensive self consistency checks among 
the data sets.  After investigating residuals of all data sets used 
in the initial fits and correlating them with the observing conditions
at the sites (seeing, sky background, and airmass), we carefully remove 
points for which we are less confident. Also, we do not use data sets
that add no or little constrain to the light curve -- such as data taken 
during only two or three nights of observations, or data taken during 
monotonic decline of the event after planetary anomalies.  The procedure 
of keeping smaller number of confident data points allows us to limit 
influence of potential systematic errors and increase our confidence in 
the results, while, due to the redundancy of the gathered data, not 
harming the discriminatory power of the light curve.

\section{Modeling The Light Curve}

Planetary lensing is a special case of the binary lensing where the mass
ratio between the lens components is very small. The description of a
binary-lensing light curve requires seven basic parameters. The first
three of these parameters characterize the geometry of the lens-source
approach. These include the time scale for the source to cross the radius 
of the Einstein ring, $t_{\rm E}$ (Einstein time scale), the time
of the closest source approach  to a reference position of the lens
system, $t_0$, and the lens-source separation at $t_0$, $u_0$ (impact 
parameter). For the reference position of the lens, we use the center of 
mass of the binary system. The Einstein ring denotes the image of a source 
for the case of the exact lens-source alignment.  Its angular radius, 
$\theta_{\rm E}$ (Einstein radius), is commonly used as a length scale 
in describing the lensing phenomenon and the lens-source impact parameter 
$u_0$ is normalized to $\theta_{\rm E}$. Another three parameters needed
to characterize the binary lens include: the mass ratio between the lens 
components, $q$, the projected binary separation in units of the Einstein 
radius, $s$, and the angle of the binary axis in respect to the lens-source 
relative motion, $\alpha$.  The last parameter is the angular source radius 
$\theta_*$ normalized to $\theta_{\rm E}$, i.e., $\rho=\theta_*/\theta_{\rm E}$ 
(normalized source radius).  This parameter is needed to describe the 
planet-induced perturbation during which the light curve is affected by 
the finite size of a source star \citep{bennett96}.

In addition to the basic binary lensing parameters, several higher-order
parameters are often needed to describe subtle light curve deviations. 
OGLE-2011-BLG-0265 lasted nearly throughout the whole Bulge season. For
such a long time-scale event, the motion of the source with respect to
the lens may deviate from a rectilinear motion due to the change of the 
observer's position caused by the Earth's orbital motion around the Sun 
and this can cause a long-term deviation in the light curve \citep{gould92a}. 
Consideration of this, so called ``parallax'' effect, in modeling a microlensing
light curve requires to  include two  additional parameters of
$\pi_{{\rm E},N}$ and $\pi_{{\rm E},E}$, which  represent the two
components of the lens parallax vector $\pivec_{\rm E}$ projected on the 
sky in the north and east equatorial coordinates, respectively. The
direction of the parallax vector corresponds to the relative
lens-source motion in the frame of the Earth at a specific time
($t_{0,{\rm par}}$). We use $t_{0,{\rm par}} = 2455760.1$. 
The size of the parallax vector is related to the Einstein radius 
$\theta_{\rm E}$ and the relative lens-source parallax  $\pi_{\rm rel} = 
{\rm AU}\,(D_{\rm L}^{-1} - D_{\rm S}^{-1})$ by
\begin{equation}
\pi_{\rm E}={\pi_{\rm rel}\over \theta_{\rm E}},
\label{eq1}
\end{equation}
where $D_{\rm L}$ and $D_{\rm S}$ are the distances to the lens and
source,  respectively. Measurement of the lens parallax is important
because it, along with the Einstein radius, allows one to determine 
the lens mass and distance to the lens as
\begin{equation}
M={\theta_{\rm E} \over \kappa \pi_{\rm E}} 
\label{eq2}
\end{equation}
and
\begin{equation}
D_{\rm L}={{\rm AU}\over \pi_{\rm E}\theta_{\rm E}+ \pi_{\rm S}},
\label{eq3}
\end{equation}
respectively \citep{gould92a}. Here $\kappa=4G/(c^2{\rm AU})$ and 
$\pi_{\rm S}={\rm AU}/D_{\rm S}$ represents the parallax of the source
star.

% Table 1 ---------------------------------------------------------------------------
\begin{deluxetable}{l|r|r}
\tablecaption{Lensing Parameters\label{table:one}}
\tablewidth{0pt}
\tablehead{parameter & \multicolumn{1}{c|}{{\it \textbf{u}}$\mathbf{_0>0}$ {\bf solution}} & \multicolumn{1}{c}{$u_0<0$ solution}}
\startdata
\multicolumn{1}{l|}{$\chi^2/{\rm dof}$} & \multicolumn{1}{r|}{4381.0/4470}  & \multicolumn{1}{r}{4386.7/4470} \\
$t_0$         (${\rm HJD}'$)         & 5760.0949 $\pm$  0.0086  & 5760.0925  $\pm$  0.0085 \\
$t_{\rm eff}$ (days)                 & 6.955     $\pm$  0.017   & -6.843     $\pm$  0.031  \\
$t_{\rm E}$   (days)                 & 53.63     $\pm$  0.19    & 53.33      $\pm$  0.27   \\
$t_*$         (days)                 & 0.5248    $\pm$  0.0055  & 0.5173     $\pm$  0.0053 \\
$q$           ($10^{-3}$)            & 3.954     $\pm$  0.063   & 3.923      $\pm$  0.059  \\
$s_0$                                & 1.03900   $\pm$  0.00086 & 1.03790    $\pm$  0.00085\\
$\alpha_0$    ($\deg$)               & -27.15    $\pm$  0.14    & 25.96      $\pm$  0.23   \\
$\pi_{{\rm E},N}$                    & 0.238     $\pm$  0.060   & 0.38       $\pm$  0.11   \\
$\pi_{{\rm E},E}$                    & 0.042     $\pm$  0.017   & 0.061      $\pm$  0.016  \\
$ds/dt$ (${\rm yr}^{-1}$)            & 0.354     $\pm$  0.019   & 0.369      $\pm$  0.019  \\
$d\alpha/dt$ ($\deg\,{\rm yr}^{-1}$) & 52.9      $\pm$  6.3     & -24.2      $\pm$  7.7    \\
$F_{\rm S,OGLE}$                     & 1.860     $\pm$  0.010   & 1.8380     $\pm$  0.0096 \\
$F_{\rm base,OGLE}$                  & 1.92436   $\pm$  0.00091 & 1.92519    $\pm$  0.00087
\enddata
\tablecomments{ 
${\rm HJD}'={\rm HJD}-2450000$. $\alpha_0$ and $s_0$ denote projected 
binary axis angle and separation for the epoch $t_{0,{\rm orb}}=2455748.0$, 
respectively.  The reference position for the definition of $t_0$ and $u_0$ 
is set as the center of mass of the lens system.  $t_{\rm eff} = u_0 \cdot 
t_{\rm E}$.  Geocentric reference frame is set in respect to the Earth 
velocity at $t_{0,{\rm par}}=2455760.1$.  Flux unit for $F_{\rm S}$ and 
$F_{\rm base}$ is 18 mag -- for the instrumental and $\sim 18.22$ for 
the calibrated OGLE {\it I}-band data.  (See Fig.~\ref{fig:three} for the 
lens geometry and Fig.~\ref{fig:five} for the CMD).}
\end{deluxetable}
%----------------------------------------------------------------------------------------

Another effect that often needs to be considered in modeling long 
time-scale lensing events is the orbital motion of the lens \citep{albrow00,
penny11, shin11, park13}. The lens orbital motion affects the light curve 
by causing both the projected binary separation $s$ and the binary axis
angle $\alpha$ to change in time.  It is especially important for the 
binary lensing systems whose separation on the sky is close to their 
Einstein ring radius (as we experience in this event).  The shape of the 
emerging ``resonant caustic'' is very sensitive to the change of the binary 
separation. Also, such caustic is considerably larger than caustics produced 
by other lens configurations allowing larger part of the lens plane to be 
accurately probed during the event.  We account for the orbital effect by
assuming  that the change rates of the projected binary separation, $ds/dt$, 
and the angular speed, $d\alpha/dt$, are constant.  This is sufficient 
approximation as we expect the orbital periods to be significantly larger 
than the 11-day period between the perturbations seen in the light curve.

Since now the binary separation is a function of time, we quote at the
tables and use as fit parameters the value of the binary separation ($s_0$)
and the binary axis angle ($\alpha_0$) for a specific epoch: $t_{0,{\rm orb}}$. 
Here we choose\footnote{Depending on the geometry of the event, different 
values of $t_{0,{\rm orb}}$ yield different correlations between parameters 
describing the event, hence, not always $t_{0,{\rm orb}}$ equal to 
$t_{0,{\rm par}}$ is the best choice in modeling.} $t_{0,{\rm orb}}$ to be 
2455748.0.  We closely follow conventions of the lensing parameters described 
in \citet{skowron11} with one difference; since we use $\alpha$ as an angle 
of the binary axis with respect to the lens-source trajectory, $d\alpha/dt$ 
describes the rotation of the binary axis in the plane of the sky.

The deviation in a lensing light curve caused by the orbital effect 
can be smooth and similar to the deviation induced by the parallax
effect. Therefore, considering the orbital effect is important as 
it might affect the lens parallax measurement and thus the physical
parameters of the lens \citep{batista11, skowron11}. 

With the lensing parameters, we test different models of the light 
curve. In the first model (standard model), the light curve is fitted 
with use of the seven basic lensing parameters. In the second model 
(parallax model), we additionally consider the parallax effect by adding 
the two parallax parameters of $\pi_{{\rm E},N}$ and $\pi_{{\rm E},E}$. 
In the third model (orbit model), we consider only the orbital motion 
of the lens by including the orbital parameters $ds/dt$ and $d\alpha/dt$, 
but do not consider the parallax effect.  In the last model 
(parallax+orbit model), we include both: the orbital motion of the lens 
and the orbital motion of the Earth (which give rise to the parallax effect).

For a basic binary model, every source trajectory has its exact mirror
counterpart with respect to the star-planet axis -- with  $(u_0, \alpha)
\rightarrow -(u_0, \alpha)$ being the only difference. However, when the 
additional effects are considered, each of the two trajectories with 
$u_0>0$ and $u_0<0$ deviate from a straight line and the pair of the 
trajectories are no longer symmetric. It is known that the models with 
$u_0>0$ and $u_0<0$ can be degenerate, especially for events associated 
with source stars located near the ecliptic plane -- this is known as 
the ``ecliptic degeneracy'' \citep{skowron11}.  For OGLE-2011-BLG-0265, the 
source star is located at $\beta\sim 2.7^\circ$ and thus we check both 
$u_0>0$ and $u_0<0$ solutions.

In modeling the OGLE-2011-BLG-0265 light curve, we search for the set
of lensing parameters  that best describes the observed light curve
by minimizing $\chi^2$ in the parameter space. We conduct this search
through three steps. In the first step,  grid searches are conducted
over the space of a set of parameters while the  remaining parameters
are searched by using a downhill approach \citep{dong06}.  We then
identify local minima in the grid-parameter space by inspecting the 
$\chi^2$ distribution. In the second step, we investigate the individual 
local minima found from the initial search and refine the individual local
solutions. In the final step, we choose a global solution by comparing 
$\chi^2$ values of the individual local minima. This multi-step 
procedure is needed to probe the existence of any possible
degenerate solutions. We choose $s$, $q$, and $\alpha$ as the grid
parameters because they are related to the light curve features in a
complex way such that a small change in their values can lead to
dramatic changes in lensing light curves. On the other hand, the light 
curve shape depends smoothly on the remaining parameters and thus 
they are searched for by using a downhill approach. For the $\chi^2$ 
minimization for refinement and characterization of the solutions, 
we use the Markov Chain Monte Carlo (MCMC) method.

% Figure 3 ---------------------------------------------------------
\begin{figure}[h]
\epsscale{1.1}
%\epsscale{0.80}
\plotone{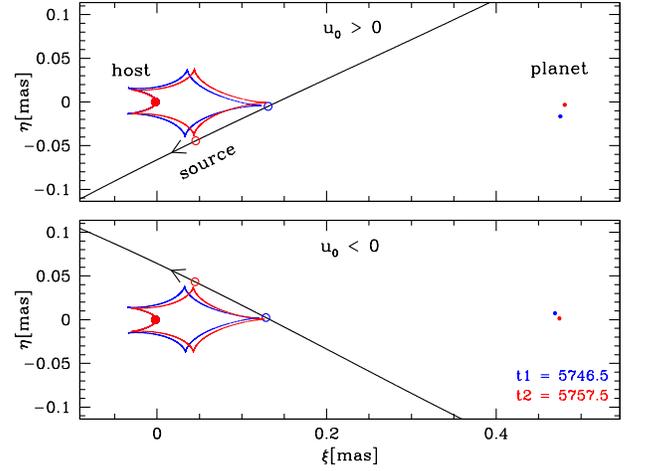}
\caption{\label{fig:three}
Geometry of the lensing system and the source star trajectory projected 
onto the plane of the sky.  The upper panel is for the best-fit
solution with $u_0>0$ and the lower panel is for the $u_0<0$ solution. 
The $u_0>0$ solution provides a slightly better fit than the $u_0<0$
solution -- by $\Delta\chi^2=5.7$.  The closed figures with cusps 
represent the caustics at two different epochs: ${\rm HJD}'=5746.5$ 
and 5757.5, which correspond to the moments of the major and minor
perturbation in the light curve. The line with an arrow represents the
source trajectory as seen from the Earth -- the curvature of the line
is due to the parallax effect. The small empty circles represent the 
size and positions of the source star at both epochs.  Also marked are 
the positions of the planet (small dots on the right) and its host star 
(big dots on the left) -- the displacement of the planet due to its 
orbital motion over 11 days between the perturbations is clearly visible. 
Origin is at the Center of Mass of the planetary system.  The horizontal 
axis is parallel with the star-planet axis at the time $t_0$.}
\end{figure}
% ----------------------------------------------------------------------

% Figure 4 ------------------------------------------------------------
\begin{figure}[th]
\epsscale{1.1}
%\epsscale{0.80}
\plotone{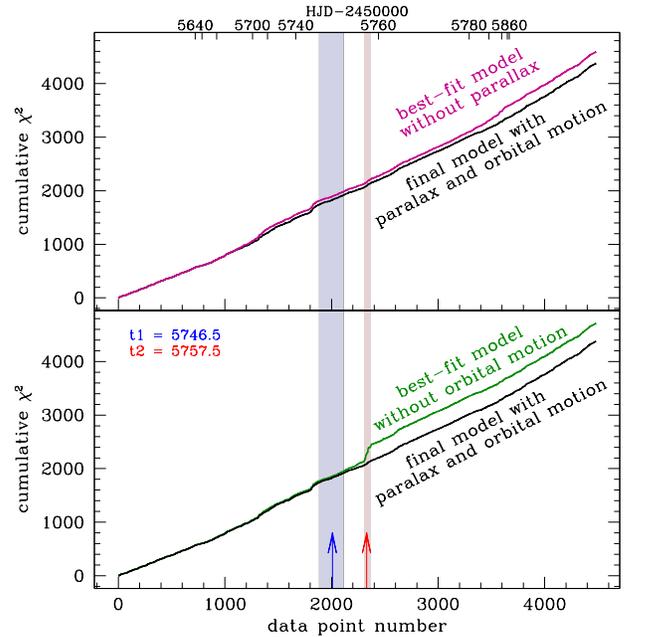}
\caption{\label{fig:four}
Cumulative $\chi^2$ distributions as a function of a data point number for 
three microlensing models. The vertical bands mark the time of the major 
and minor planetary anomaly. We see, that the minor anomaly cannot be a 
well fitted without the inclusion of the lens orbital motion.  Two anomalies 
appeared in the light curve closer in time than could be fitted with a static 
binary model. Since the first anomaly is extremely well covered with observations, 
it is the second anomaly that does not fit the best possible static model.
}\end{figure}
% -----------------------------------------------------------------------

A planetary perturbation is mostly produced by the approach of the source
star close to caustics that represent the positions on the source plane 
at which the lensing magnification of a point source becomes infinite.  
During the approach, lensing magnifications are affected by finite-source 
effects due to the differential magnification caused by the steep gradient 
of magnification pattern around the caustic. For the computation of 
finite-source magnifications, we use the ray-shooting method
\citep{schneider86,kayser86, wambsganss97}. In this method, a large
number of rays are uniformly shot from the image plane, bent according
to the lens equation, and land on the source plane. The lens equation 
for image mapping from the image plane to the source plane is expressed as 
\begin{equation}
\zeta=z-{1\over 1+q} \left( {1\over \bar{z}-\bar{z}_{{\rm L},1} } +
{q\over \bar{z}-\bar{z}_{{\rm L},2}} \right),
\label{eq4}
\end{equation}
where $\zeta$, $z_{\rm L}$, and $z$ are the complex notations of the
source, lens  and image positions, respectively, and the overbar denotes
complex conjugate.   Here all lengths are expressed in units of the
Einstein radius.  The finite  magnification is computed as the ratio of
the number  density of rays on the source surface to the density on the
image plane. This numerical  technique requires heavy computation and
thus we limit finite-magnification computation  based on the ray-shooting 
method to the region very close to caustics. In the adjacent region, we 
use a hexadecapole approximation, with which finite magnification  
computation can be faster by several orders of magnitude \citep{pejcha09, 
gould08}.  We solve lens equation by using the complex polynomial method 
described in \citet{skowron12}.

In computing finite-source magnifications, we incorporate the
limb-darkening variation of the stellar surface brightness. The surface
brightness profile is modeled as $S_\lambda \propto 1 - c_\lambda \cdot 
(1-\cos\phi) - d_\lambda \cdot (1-\sqrt{\cos\phi})$, where $c_\lambda$ and 
$d_\lambda$ are the limb-darkening coefficients of the wavelength band 
$\lambda$ and $\phi$ is the angle between the normal to the stellar surface 
and the line of sight toward the center of the star.  Based on the stellar 
type (see Section 5), we adopt the coefficients using Table 32 (square-root 
law) of \citet{claret00} for $v_{\rm t} = 2$, solar metallicity, 
$T_{\rm eff}=5000\,{\rm K}$ and $\log g = 3.5$:
\begin{align}
  c_{I-band},\, d_{I-band} &= 0.2288,\, 0.4769, \\
  c_{MOA-R}, \, d_{MOA-R}  &= 0.2706,\, 0.4578, \\
  c_{V-band},\, d_{V-band} &= 0.5337,\, 0.2993.
\end{align}
Here the values for the non-standard MOA-{\it R} filter are taken as linear 
combination of {\it R}-band and {\it I}-band coefficients with
30\% and 70\% weights.

\section{Results}

\subsection{Best-fit Solution}
\label{sec:solution}

In Table~\ref{table:one}, we present the best-fit solutions along with 
their $\chi^2$ values.  In order to provide information about the blended 
light (i.e., the light that was not magnified during the event),
we also present the source, $F_{\rm S}$, and baseline, $F_{\rm base}$, 
fluxes estimated from the OGLE photometry.  We note that the uncertainty 
of each parameter is estimated based on the distribution in the MCMC 
chain obtained from modeling.

It is found that the perturbation was produced by a planetary companion
with a  planet/star mass ratio $q\sim 3.9\times 10^{-3}$ located close
to the Einstein  ring of its host star, i.e, $s\sim 1.0$. In the upper
panel of Figure~\ref{fig:three}, we present the locations of the lens 
components, the caustic, and the source trajectory for the best-fit solution. 
Since the planet is close to the Einstein ring, the resulting caustic forms 
a single closed curve with six cusps. It is found that the major (at 
${\rm HJD}'=5746.5$) and minor (at ${\rm HJD}'=5757.5$) perturbations in 
the lensing light curve were produced by the approach of the source star 
close to the strong and weak cusps of the caustic, respectively.

We find that the event suffers from the ecliptic degeneracy. In
Figure~\ref{fig:three}, we compare  the lens-system geometry of the two
degenerate solutions with $u_0<0$ and $u_0>0$.  We note that the source
trajectories of the two degenerate solutions are almost symmetric with
respect to the star-planet axis. The $\chi^2$ difference between the two 
degenerate models is merely 5.7 -- with $u_0>0$ solution slightly preferred 
over the $u_0<0$ solution. We further discuss this degeneracy in Section 5.2.

Higher-order effects are important for the event. We find that the model
considering the parallax effect improves the fit with $\Delta\chi^2 = 230.9$ 
for $u_0>0$  and 127.8 for $u_0<0$ compared to the standard model.  The model 
considering the lens orbital motion (but without parallax) improves the fit 
even more with $\Delta\chi^2 = 349.1$ compared to the standard model.  
Considering both the parallax and orbital effects yields a light curve model 
that fits the data significantly better with $\Delta\chi^2 = 559.4$ for 
$u_0>0$  and 565.1 for $u_0<0$ relative to the standard model.

The importance of the lens orbital motion can be seen in 
Figure~\ref{fig:four}. It shows cumulative $\chi^2$ distribution for 
the full (final) model and compares it to the models without the parallax 
effect (upper panel) and without the lens orbital motion (lower panel)
taken into account.  It is found that the signal of the orbital effect 
is mainly seen from the part of the light curve at around ${\rm HJD}'\sim
5757.5$, which corresponds to the time of the minor anomaly.  The second 
anomaly happened sooner than predicted by the static binary model. Without 
the observations at this time, we would have lacked the information on the 
evolution of the the caustic shape during the time between the anomalies.
The minor anomaly was densely covered by follow-up data, especially the 
SAAO data,  but the coverage by the survey data is sparse. As a result, 
the orbital parameters could not be well constrained by the survey data 
alone.

% Figure 5 -----------------------------------------------------
\begin{figure}[th]
\epsscale{1.1}
%\epsscale{0.85}
\plotone{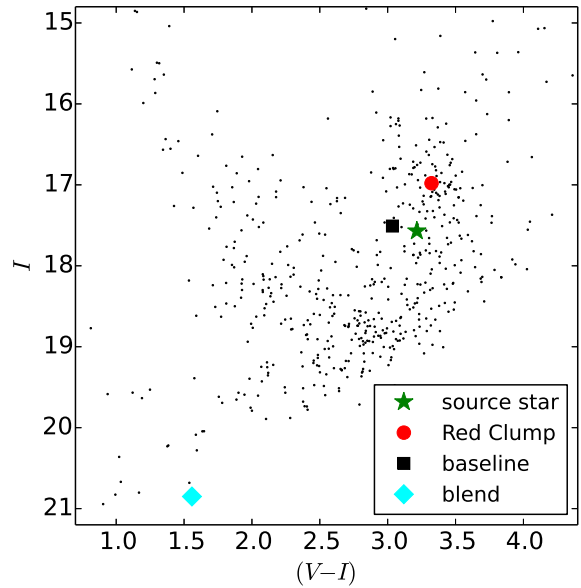}
\caption{\label{fig:five}
Location of the source star in the OGLE color-magnitude diagram for stars 
in the $1.5'\times 1.5'$ field centered on the lensing event. Also shown 
is the centroid of the red clump giant stars that is used to calibrate 
the brightness and color of the  source star.  Light seen before and 
after the event (baseline level) was separated into the ``source''  and 
the ``blend'' light (or the portion of light that was magnified and the 
portion that was not magnified during the course of the event, respectively)}.
\end{figure}
% ---------------------------------------------------------------

\subsection{Angular Einstein radius}
\label{sec:theta_star}

Detection of the microlens parallax enables the measurement of the mass 
and distance to the planetary system.  The Einstein radius, the
second component required by Equations~(\ref{eq2}) and (\ref{eq3}), is
estimated by $\theta_{\rm E}=\theta_*/\rho$, where the angular source
radius $\theta_*$ is obtained from the color and brightness information
and the normalized source radius $\rho$ is measured from the
microlensing light curve fitting to the planetary perturbation.

\subsubsection{Intrinsic color and extinction-corrected brightness of the source star}

To determine the angular source radius, we first locate the source star 
on the color-magnitude diagram for stars in the field and then calibrate
its de-reddened color and brightness by using the centroid of the red
clump giants as a reference under the assumption that the source and red
clump giants experience the same amount of extinction and reddening
\citep{yoo04}.

In Figure~\ref{fig:five}, we present the location of the source star in
the color-magnitude diagram.  Using the method of \citet{nataf10} finding 
the centroid of the red clump in the 1.5' x 1.5' region of the sky around 
the source star, we estimate that the source star (with $V=20.77$ and 
$I=17.57$) is $0.12$ mag bluer and $0.59$ mag fainter than the typical 
red clump giant, and hence, is most likely also a K-type giant star located 
in the Galactic bulge. Based on the intrinsic color of the red clump giant 
stars $(V-I)_{{\rm RC}, 0} = 1.06$ \citep{bensby11}, we estimate the 
de-reddened color of the source star to be $(V-I)_{{\rm S}, 0} = 0.94$.

With the observed ($(V-I)_{\rm RC} = 3.32$) and intrinsic colors of the 
red clump stars, we estimate the total reddening toward the Galactic Bulge:
\begin{equation}
  E(V-I) = (V-I)_{\rm RC} - (V-I)_{{\rm RC}, 0} = 3.32 - 1.06 = 2.26.
\end{equation}
From \cite{nataf13} the mean distance to the Galactic bulge stars in the 
direction of the event is 7.8 kpc and the intrinsic brightness of the red 
clump stars is $M_{I, {\rm RC}, 0} = -0.11$.  With the measured observed 
brightness $I_{\rm RC} = 16.98$, we estimate the extinction to the bulge
to be $A_I = 2.63$.  This is consistent with the estimated reddening
of 2.26 as the slope of the reddening vector ($\partial A_I/\partial E(V-I)$) 
is typically $\sim 1.2$, and in most cases is between 1.0 and 1.4 for the 
Galactic bulge sight lines \citep[Figure 7]{nataf13}.  The extinction in 
the $V$ band is calculated as $A_V = A_I + E(V-I) = 4.89$.  Then, the 
extinction-corrected magnitudes of the source star are computed as
\begin{align}
  V_{{\rm S}, 0} &= V_{\rm S} - A_V = 20.77 - 4.89 = 15.88, \\
  I_{{\rm S}, 0} &= I_{\rm S} - A_I = 17.57 - 2.63 = 14.94.
\end{align}

\subsubsection{Uncertainties of the source color estimation}

Although uncertainties of the observed color and brightness of the 
source stars are typically low (in this case 0.01 mag), the uncertainty 
in the centroiding of the red clump and the differential reddening in 
the field causes that the true intrinsic colors of the microlensing 
sources are typically known with lower accuracy.  
\citet[Section 3.2]{bensby13} compare the colors of source stars of 
the 55 microlensing events determined with both spectroscopic and 
microlensing techniques.  Their Figure 5 shows that the disagreement 
between the two estimations is typically 0.07 mag for the blue star 
sample and 0.08 mag for all stars.  There is no physical reason for 
the measurement of the color offset from the red clump stars to be less 
accurate for red stars than for blue stars. Hence, the authors point to, 
clearly but not perfectly, the color-$T_{\rm eff}$ relations as the 
source of the increased scatter for red stars (with $T_{\rm eff}<5500$ K, 
cf.  \citealt{bensby13}, Figure 7).  The observed 0.07 mag scatter between 
the spectroscopic and microlensing color estimates also includes the 
uncertainties in $T_{\rm eff}$, which are of the order of $100$ K and 
would generate $\sim 0.034$ mag uncertainty in color (compare with 
Table 5 and Figure 7 of \citealt{bensby13}).  By subtracting this source 
of scatter in quadrature from the observed scatter, we obtain 0.061 mag, 
which still contains some unknown uncertainty of the stellar models 
themselves.

The sample of events analyzed by \citet{bensby13} also contains
some problematic events of two types. One type are the events where
the coverage of the light curve in the multiple photometric bands 
was not sufficient to accurately determine observed color, while the 
other type are the events in the fields with poorly defined red clumps.
This allows us to argue that for well observed microlensing events in 
the fields with well defined red clump, the typical error in the 
microlensing color estimation is on the order of $0.05$ mag.

One could worry that the assumption of a typical error of the intrinsic 
color estimation does not take into account the influence of the 
differential reddening, which in fact, varies from field to field. 
Figure 6 of \citet{bensby13} addresses this issue, showing that there 
is no evidence of strong correlation between the differential reddening 
in the fields of 55 events (as measured by \citet{nataf13}) and errors 
in their color estimations.  This actually could be understood by 
realizing that the dominating source of scatter in the observed colors 
of red clump stars comes from the gradient of the reddening across 
the field. This gradient, however, has no effect on the position of 
the red clump stars centroid.

As an example, we took samples of stars from four circles 
centered on our event with the diameters of 1.5', 2', 3', and 4', 
respectively, and used them to measure the centroid of the red clump. 
All four measurements are within 0.02 mag of each other, even though the 
measure of the differential reddening (as defined by \citealt{nataf13}) 
is between 0.16 and 0.24 in these circles.

It is also worth noting that any error in the relative position of 
the source star from the centroid of the red clump  that could come 
from the differential reddening would only partially contribute to 
the final estimation of the angular size of the star. As we will see 
in the following section, the calibration of the angular radius of the 
star contains two terms with opposite signs: $\sim0.5 (V-I)_0 - 0.2 I_0$ 
(Eq. (\ref{eq:kf08})). More dust in front of the star influences both
estimations of $(V-I)_0$ and $I_0$ in the same direction, and since
$A_I \approx 1.2\, E(V-I)$, the overall error that comes from the
wrong estimation of the reddening is reduced by 50\%
($1.2 \times 0.2/0.5 = 0.5$).

The OGLE-2011-BLG-0265 event is located $1.5 \deg$ from the Galactic 
plane in the region strongly obscured by the interstellar dust 
($E(V-I)=2.26$) and affected by the differential reddening. Following 
\citet{nataf13}, we calculate the measure of the differential reddening 
($\sigma_{E(V-I)}$) in the $1.5' \times 1.5'$ patch of the sky around 
the event.  The observed $V-I$ colors of the red clump stars show 
$0.20$ mag dispersion, which leads to the estimation of $\sigma_{E(V-I)}
= 0.16$ mag.  However, having the evidence for at most minor influence 
of the differential reddening on the final estimation of the color,
and knowing that only half of the error (due to reddening) enters
the final result, we only slightly increase our uncertainty of 
the color from 0.05 to 0.06 mag due to the heavily reddened field.

We expect the error in the estimation of $I_0$ to 
be slightly higher than the assumed error for $(V-I)_0$. In order 
to measure the observed brightness of the red clump, the luminosity 
function of the red giant branch has to be fitted simultaneously 
with the luminosity function of the red clump giants. 
Based on the reproducibility of the red clump centroiding under 
various assumptions regarding the red giant branch luminosity function, 
we conservatively assume 0.1 magnitude error in the estimation of $I_0$
of the source star.

\subsubsection{Angular size from the surface brightness relations}

Knowing the dereddened color of the star and the extinction-corrected 
brightness enables the use of the surface brightness relation to find 
the angular radius ($\theta_*$).  We note that in microlensing we 
typically measure $(V-I)_0$ color, hence, ideally we would like to use
a calibration based on this quantity.  By including the additional 
transformation process from $(V-I)_0$ to $(V-K)_0$, the uncertainty 
of the estimated color increases by a factor 1.5 -- 2.5 (for example 
$\partial (V-K)/\partial (V-I)\approx 2.5$ for stars with $(V-I)<1.3$
\citep{bessell88}).  \cite{kervella-fouque08} provide such a relation
calibrated with dwarfs and subgiant stars; we write:
\begin{equation}
  \log \theta_* = 3.1982 - 0.2 I_0 + 0.4895 (V-I)_0 - 0.0657 (V-I)_0^2, 
  \label{eq:kf08}
\end{equation}
where the angular radius is given in $\mu$as and 
the scatter of the relation is 0.0238.
The relation in $(V-K)$ for the same types of stars based on
\cite{kervella04a} is
\begin{equation}
  \log \theta_* = 3.2165 - 0.2 V_0 + 0.2753 (V-K)_0 
\end{equation}
and the quadratic 
relation for the wider range of stars was given by \cite{dibenedetto05}:
\begin{equation}
  \log \theta_* =  3.2120 - 0.2 V_0 + 0.2968 (V-K)_0 - 0.0088 (V-K)_0^2. 
\end{equation}

\citet{kervella-fouque08} believe that the scatter around the provided 
relation is dominated by the intrinsic scatter rather than measurement
errors.  This yields the relative uncertainty of the angular radius at 
5.5\%.  Calibrations based on an infrared color have much smaller intrinsic 
scatter, so careful removal of scatter due to measurement error is required.
\citet{kervella04a} estimate the intrisic scatter around the provided
relation is 1\%, whereas \cite{dibenedetto05} estimates 1.8\% and argues 
that the accuracy of the star sizes obtained from the infrared-based 
surface brightness relations is $< 2\%$, but higher than the 1\%
estimated by \citet{kervella04a}.

We note that despite having much smaller scatter, the relations with 
$(V-K)_0$ (transformed from $(V-I)_0$) yield higher uncertainty of the 
angular radius than the relation originally calibrated in $(V-I)_0$, 
unless the accuracy of $(V-I)_0$ estimation is $\lesssim 0.05$ or 
$(V-I)_0>1.3$, where the slope of $(V-K)$ vs $(V-I)$ is more shallow.
Hence, we use \cite{kervella-fouque08} relation in the OGLE-2011-BLG-0265 
case, which leads to the final estimation of the angular source radius
\begin{equation}
  \theta_*= 4.09 \pm 0.41\ \mu{\rm as}.
  \label{eq:theta_star}
\end{equation}

Combining the physical and the normalized source radius yields the 
Einstein radius of $\theta_{\rm E} = \theta_*/\rho = 0.42\pm 0.04$ mas 
(for both $u_0>0$ and $u_0<0$ solutions of the parallax+orbit model).

\subsection{Physical Parameters}

With the measured Einstein radius and the lens parallax, we are able 
to estimate the physical quantities of the lens system 
(Table~\ref{table:two}). For the best-fit solution ($u_0>0$, 
parallax+orbit model), the lens mass and distance to the lens are
$M=0.22\pm 0.06\ M_\odot$ and $D_{\rm L}=4.4\pm 0.5$ kpc, respectively. 
The mass of the planet is $M_{\rm p}=0.9\pm 0.3\ M_{\rm J}$.  The 
projected separation between the host star and the planet is $a_\perp=
1.9\pm 0.2$ AU and thus the planet is located well beyond the snow line 
of the host star.

For the marginally disfavored $u_0<0$ solution, the resulting 
physical parameters of the lens system are somewhat different --
as expected, mainly caused by 
the difference in the north component of the parallax vector, i.e., 
$\pi_{{\rm E},N}$. See Table~\ref{table:one}. In this case, the lens 
mass and distance are $M=0.14\pm 0.06\ M_\odot$ and $D_{\rm L}=
3.5\pm 0.7$ kpc, respectively, and the mass of the planet is 
$M_{\rm p}=0.6\pm 0.3\ M_{\rm J}$.

Hence, the system belongs to a little-known population of planetary 
systems where a Jupiter mass planet orbits an M~dwarf beyond its 
snow line.

% Table 2 ---------------------------------------------------------------------------------------------
\begin{deluxetable}{lcccc}
\tablecaption{Physical Lens Parameters\label{table:two}}
\tablewidth{0pt}
\tablehead{
\multicolumn{1}{l}{quantity} &
  \multicolumn{1}{c}{{\it \textbf{u}}$\mathbf{_0>0}$ {\bf solution}} &
\multicolumn{1}{c}{$u_0<0$ solution}
}
\startdata
$\theta_{\rm E}$ (mas)       & $0.419 \pm 0.040$           & $0.423 \pm 0.040$            \\
$\mu_{\rm geo}$ (mas/yr)     & $2.85 \pm 0.29$             & $2.89 \pm 0.29$              \\
$\mu_{{\rm hel}, N}$ (mas/yr)& $2.72 \pm 0.30$             & $2.73 \pm 0.32$              \\
$\mu_{{\rm hel}, E}$ (mas/yr)& $1.06_{-0.18}^{+0.29}$      & $1.39_{-0.16}^{+0.19}$       \\
$M_{\rm p}$ ($M_{\rm J}$)    & $0.88_{-0.18}^{+0.27}$      & $0.56_{-0.13}^{+0.25}$       \\
$M_{\rm h}$ ($M_\odot$)      & $0.211_{-0.045}^{+0.068}$   & $0.136_{-0.031}^{+0.061}$    \\
$D_{\rm L}$ (kpc)            & $4.38_{-0.45}^{+0.51}$      & $3.49_{-0.49}^{+0.71}$       \\
$a_\perp$ (AU)               & $1.89_{-0.20}^{+0.25}$      & $1.51_{-0.20}^{+0.31}$       \\
${\rm (KE/PE)}_\perp$        & $0.387 \pm 0.075$           & $0.112 \pm 0.060$                              
\enddata
\tablecomments{
Parameters calculated for parallax+orbital model;
$\theta_{\rm E}$: angular Einstein radius, $\mu_{\rm geo}$ and $\mu_{\rm
hel}$:  relative lens-source proper motion in the geocentric and
heliocentric reference frames, respectively, $M_{\rm p}$: mass of the
planet, $M_{\rm h}$: mass of the host star, $D_{\rm L}$: distance to the
lens, $a_\perp$: projected star-planet, ${\rm (KE/PE)}_\perp$: the ratio
of the transverse kinetic to potential energy.
}
\end{deluxetable}
% -----------------------------------------------------------------------------------------------------

\section{Discussion}

\subsection{Degeneracy of the Microlensing Models}

While the planetary nature of the perturbation in the OGLE-2011-BLG-0265
light curve is obvious, the event suffers from the orbiting binary ecliptic 
degeneracy \citep[see Appendix 3]{skowron11}.  The two solutions, $u_0<0$ 
and $u_0>0$, have nearly identical mass ratio $q=3.9\times 10^{-3}$, 
normalized separation $s=1.04$, Einstein radius $\theta_{\rm E}=0.42\,$ 
mas, and hence planet-host angular separation $\theta_\perp=0.44\,$ mas. 
They differ in the microlens parallax, especially in the north component 
$\pi_{{\rm E},N} = 0.24$ ($u_0>0$) versus $\pi_{{\rm E},N} = 0.38$ ($u_0<0$), 
and also in $d\alpha/dt$, which is often strongly correlated with 
$\pi_{{\rm E},N}$ \citep{batista11, skowron11}. This difference is 
important because it leads to a different mass and distance of the host 
$(M_*/M_\odot,D_{\rm L}/{\rm kpc})=(0.22,\, 4.4)$ versus (0.14, 3.5) for 
the $u_0>0$ and $u_0<0$ solutions, respectively.

The $u_0>0$ solution is favored by $\Delta\chi^2=5.7$ corresponding to 
a (frequentist) likelihood ratio of $\exp(5.7/2)=17$.  This would be 
compelling evidence if treated at face value. However, it is known that 
the photometry of microlensing events occasionally suffers from low-level 
systematic trends at $\Delta\chi^2\sim$ few level.

As an additional way to resolve the degeneracy, we check the ratio 
of the projected kinetic to potential energy \citep{dong09} estimated by
\begin{equation}
\beta = \biggl({\rm KE\over PE}\biggr)_\perp =
{v_\perp^2 r_\perp\over 2G M_{\rm tot}}
= {\kappa M_\odot{\rm yr}^2\over 8\pi^2}{\pi_{\rm E} s^3\gamma^2\over 
\theta_{\rm E}(\pi_{\rm E} + \pi_S/\theta_{\rm E})^3},
\label{eq5}
\end{equation}
where $\gamma^2=(ds/dt/s)^2 + (d\alpha/dt)^2$.  For typical viewing
angles, one expects $\beta\sim {\cal O}(0.4)$, as is the case for the
$u_0>0$ solution. On the other hand, the lower value $\beta\sim 0.11$ 
(as in the $u_0<0$ solution) implies either that the planet is seen 
projected along the line of sight at the viewing angle $\psi\sim 2
\beta\sim 0.22$ (corresponding to semi-major axis $a\sim a_\perp/(2\beta)
\sim 4.5 a_\perp$), or that we have just seen the planet when majority of
its motion is directly toward (or away from) us. Of course, the prior
probability of the first configuration for a point randomly distributed
on a sphere is just $2\beta^2\sim 0.025$ and the probability
for the second is similar.

However, while it is certainly true that the prior probability for any
given planet's position is uniform over a sphere, this is not the case
for planets found by microlensing, which are preferentially detected
within a factor $\sim 1.5$ of the Einstein radius \citep{gould92b, gould10}.  
First, since the planet actually lies very near the Einstein radius, it
would have been detected in almost any angular orientation $\alpha$, so
that the actual probability is more like $\beta$ than $\beta^2$, i.e.,
about 0.11.  In addition, since giant planets around M~dwarfs are a new 
class of planets, we do not know their distribution. It could be that 
the great majority of such planets lie at $a\sim 20\,$AU. Then, whenever
these were found in planet-host microlensing events, they would have 
a very low value $\beta$.  On the other hand, whenever we detect them 
at typical viewing angles, they will be considered as ``free floating 
planets'' \citep{sumi11}.  Hence, the measurement of a low $\beta$ value 
is not a strong statistical argument against the $u_0<0$ solution which 
still remains as a viable option.

In summary, although both light curve and energy considerations point 
to the $u_0>0$ solution, it is difficult to confidently resolve the 
degeneracy between the two possible models based on the currently 
available data.  Fortunately, the difference in the physical parameters 
estimated from the two degenerate solutions are not big enough to affect
the conclusion that the lens belongs to a new class of giant planets 
around low-mass stars.

\subsection{Prospects for Follow-up Observations}

It will eventually be possible to confidently resolve the current 
degeneracy issue on the models of OGLE-2011-BLG-0265 when the Giant
Magellan Telescope (GMT) comes on line in about 10 years. At that time,
the source and lens will be separated by about 40 mas, or roughly 3
FWHM in the $J$-band. There are three observables that can then be used
to distinguish the two solutions. First, the $u_0<0$ solution predicts
a fainter lens because it has a lower mass.  Second, it predicts slightly 
higher heliocentric proper motion (mainly in the East sky direction). 
Third, it predicts a different angle of the proper motion.

Each of these measurements has some potential problems. The prediction 
of the lens flux is influenced not only by the mass and the distance 
but also by the extinction to the lens.  There is a substantial error 
in $\theta_{\rm E}$ that impacts the brightness in the same direction
as the mass and distance.  That is, if $\theta_{\rm E}$ is higher than 
we have estimated, then the distance is closer (so the lens is brighter) 
and the mass is greater (so the lens is brighter again).  Fortunately, 
the detection of the lens will itself enable to measure the proper motion 
$\mu_{\rm hel}$ and therefore also the Einstein radius (see below).  
Also, with a bigger telescope, it will be possible to better estimate 
$\theta_{\rm E}$ by detailed characterization of the source star, thus, 
more accurate estimation of the $\theta_*$ then in Eq (\ref{eq:theta_star}).

As can be seen from Table~\ref{table:two}, the predictions for heliocentric
proper motion differ by only $1\sigma$.  This is the same problem as just 
mentioned: the proper motion prediction contains the significant error of 
$\theta_{\rm E}$.

By contrast, the angle of heliocentric proper motion, $\phi_\mu$, does
not depend on $\theta_{\rm E}$.  In terms of observables,
\begin{equation}
\muvec_{\rm hel} = {\theta_{\rm E}\over t_{\rm E}}\,{\pivec_{\rm E}
\over \pi_{\rm E}} + 
{{\bf v}_{\oplus,\perp}\over {\rm AU}}\pi_{\rm rel}
= \theta_{\rm E}\biggl({1\over t_{\rm E}}{\pivec_{\rm E}\over \pi_{\rm E}} +
{{\bf v}_{\oplus,\perp}\over {\rm AU}}\pi_{\rm E}\biggr),
\label{eq6}
\end{equation}
where ${\bf v}_{\oplus,\perp}$ is the motion of Earth projected on the
sky at the fiducial time of the event  $(v_{\oplus,N},v_{\oplus,E}) =
(-0.42,5.45)\,$AU/yr.  This means that the position angle (North
through East) is
\begin{equation}
\phi_\mu = {\rm atan}{\pi_{{\rm E},E} + \pi_{\rm E}^2 v_{\oplus,E} 
t_{\rm E}/{\rm AU}
\over \pi_{{\rm E},N} + \pi_{\rm E}^2 v_{\oplus,N} t_{\rm E}/{\rm AU}},
\label{eq7}
\end{equation}
which is indeed independent of $\theta_{\rm E}$. We find $\phi_\mu =
20.8_{-2.7}^{+5.3} \deg$ for $u_0>0$ and $\phi_\mu = 26.4_{-1.2}^{+2.0} 
\deg$ for $u_0<0$.  Thus if the actual measurement is
$\phi_\mu=21^\circ$, it will strongly exclude $(4\,\sigma)$ the $u_0<0$
solution, but if it is $\phi_\mu=26^\circ$ then it will only marginally
favor $(1\,\sigma)$ the $u_0<0$ solution.

Nevertheless, with three pieces of
information, there is a good chance that the ensemble of measurements
will favor one solution or the other.

\section{Conclusions}

We reported the discovery of a planet detected by analyzing the light
curve of the microlensing event OGLE-2011-BLG-0265. It is found  that
the lens is composed of a giant planet orbiting a M-type dwarf
host. Unfortunately, the microlensing modeling yields two degenerate 
solutions, which increase our uncertainties in mass of and distance to
this planetary system and cannot be distinguished with currently 
available data. Planet-host mass ratio is, however, very well measured 
at $0.0039$.

The slightly preferred solution yields a Jupiter-mass planet orbiting 
a $0.22 M_\odot$ dwarf.  The second solution yields a 0.6 Jupiter-mass 
planet orbiting a $0.14 M_\odot$ dwarf. There are good prospects for
lifting the degeneracy of the solutions with future additional
follow-up observations. In either case,  OGLE-2011-BLG-0265 event
demonstrates the uniqueness of the microlensing  method in detecting
planets around low-mass stars.

\acknowledgments

The OGLE project has received funding from the European
Research Council under the European Community's Seventh
Framework Programme (FP7/2007-2013) / ERC grant agreement
no. 246678 to A.~Udalski. 
This research was partly supported by the Polish Ministry of Science
and Higher Education (MNiSW) through the program ``Iuventus Plus''
award No. IP2011 026771.
Work by C.~Han was supported by Creative Research Initiative Program
(2009-0081561) of National Research Foundation of Korea. 
The MOA experiment was supported by grants JSPS22403003 and
JSPS23340064. T.S. acknowledges the support JSPS 24253004. T.S. is
supported by the grant JSPS23340044. Y.M. acknowledges support from 
JSPS grants JSPS23540339 and JSPS19340058.
Work by J.C.~Yee is supported in part by a Distinguished University
Fellowship from The Ohio State University and in part under contract
with the California Institute of Technology (Caltech) funded by NASA
through the Sagan Fellowship Program.
Work by A.G. and B.S.G was supported by NSF grant AST 1103471. Work 
by A.G., B.S.G., and RWP was supported by NASA grant NNX12AB99G. TS
acknowledges the support from the grant JSPS23340044 and JSPS24253004.
CS received funding from the European Union Seventh Framework Programme
(FP7/2007-2013) under grant agreement no. 268421.
This work is based in part on data collected by MiNDSTEp with the 
Danish 1.54m telescope at the ESO La Silla Observatory. The Danish 
1.54m telescope is operated based on a grant from the Danish Natural 
Science Foundation (FNU). The MiNDSTEp monitoring campaign is powered 
by ARTEMiS (Automated Terrestrial Exoplanet Microlensing Search; 
Dominik et al. 2008, AN 329, 248). MH acknowledges support by the 
German Research Foundation (DFG). DR (boursier FRIA) and JSurdej 
acknowledge support from the Communaut\'{e} fran\c{c}aise de Belgique 
-- Actions de recherche concert\'{e}es -- Acad\'{e}mie universitaire 
Wallonie-Europe. KA, DMB, MD, KH, MH, CL, CS, RAS, and YT are thankful 
to Qatar National Research Fund (QNRF), member of Qatar Foundation, 
for support by grant NPRP 09-476-1-078.
Work by D.D.P. was supported by the University of Rijeka Project 
13.12.1.3.02.
This research was supported by the I-CORE program of the Planning 
and Budgeting Committee and the Israel Science Foundation, Grant 
1829/12.  DM and AG acknowledge support by the US-Israel Binational 
Science Foundation.


\begin{thebibliography}{99}

\bibitem[Alard \& Lupton(1998)]{alard98}
Alard, C., \& Lupton, R.~H.\ 1998, \apj, 503, 325

\bibitem[Albrow et al.(2000)]{albrow00}
Albrow, M.~D., Beaulieu, J.-P., Caldwell, J.~A.~R., et al. 2000, \apj, 534, 894

\bibitem[Albrow et al.(2009)]{albrow09}
Albrow, M.~D., Horne, K., Bramich, D.~M., et al. 2009, \mnras, 397, 2099

\bibitem[Batista et al.(2011)]{batista11}
Batista, V., Gould, A., Dieters, S., et al. 2011, \aap, 529, A102

\bibitem[Beaulieu et al.(2006)]{beaulieu06}
Beaulieu, J.-P., Bennett, D.~P., Fouqu\'e, P., et al. 2006, Nature, 439, 437

\bibitem[Bennett \& Rhie(1996)]{bennett96}
Bennett, D.~P., \& Rhie, S.~H.\ 1996, \apj, 472, 660

\bibitem[Bennett et al.(2008)]{bennett08}
Bennett, D.~P., Bond, I.~A., Udalski, A., et al. 2008, \apj, 684, 663

\bibitem[Bennett et al.(2010)]{bennett10}
Bennett, D.~P., Rhie, S.~H., Nikolaev, S., et al. 2010, \apj, 713, 837

\bibitem[Bensby et al.(2011)]{bensby11} 
Bensby, T., Ad{\'e}n, D., Mel{\'e}ndez, J., et al.\ 2011, \aap, 533, AA134 

\bibitem[Bensby et al.(2013)]{bensby13}
Bensby, T. Yee, J.C., Feltzing, S. et al. 2013, \aap, 549A, 147

\bibitem[Bessell \& Brett(1988)]{bessell88}
Bessell, M.\ S., \& Brett, J.\ M.\ 1988, \pasp, 100, 1134

\bibitem[Bond et al.(2001)]{bond01}
Bond, I.~A., Abe, F., Dodd, R.~J., et al. 2001, \mnras, 327, 868

\bibitem[Bonfils et al.(2011)]{bonfils11}
Bonfils, X., Gillon, M., Forveille, T., et al. 2011, \aap, 528, 111

\bibitem[Borucki et al.(2011)]{borucki11}
Borucki, W. J., Koch, D. G., Basri, G., et al. 2011, \apj, 736, 19

\bibitem[Boss(2006)]{boss06}
Boss, A.~P.\ 2006, \apj, 643, 501

\bibitem[Bramich(2008)]{bramich08}
Bramich, D.~M.\ 2008, \mnras, 386, L77

\bibitem[Cassan et al.(2012)]{cassan12}
Cassan, A., Kubas, D., Beaulieu, J.-P., et al. 2012, Nature, 481, 167

\bibitem[Charbonneau et al.(2009)]{charbonneau09}
Charbonneau, D., Berta, Z. K., \& Irwin, J. 2009, Nature, 462, 891

\bibitem[Claret(2000)]{claret00}
Claret, A.\ 2000, \aap, 363, 1081

\bibitem[Delfosse et al.(1998)]{delfosse98}
Delfosse, X., Forveille, T., Mayor, M., Perrier, C., 
Naef, D., \& Queloz, D.\ 1998, \aap, 338, L67

\bibitem[Di Benedetto(2005)]{dibenedetto05} Di Benedetto, G.~P.\
2005, \mnras, 357, 174

\bibitem[Dominik et al.(2010)]{dominik10} Dominik, M., 
J{\o}rgensen, U.~G., Rattenbury, N.~J., et al.\ 2010, Astronomische 
Nachrichten, 331, 671 

\bibitem[Dong et al.(2006)]{dong06}
Dong, S., DePoy, D.~L., Gaudi, B.~S., et al. 2006, \apj, 642, 842

\bibitem[Dong et al.(2009)]{dong09}
Dong, S., Gould, A., Udalski, A., et al. 2009, \apj, 695, 970

\bibitem[Gaudi et al.(2008)]{gaudi08}
Gaudi, B.~S., Bennett, D.~P., Udalski, A., et al. 2008, Science, 319, 927

\bibitem[Gillon et al.(2007)]{gillon07}
Gillon, M., Pont, F., Demory, B.-O., et al. 2007, \aap, 472, L13

\bibitem[Gould(1992)]{gould92a}
Gould, A.\ 1992, \apj, 392, 442

\bibitem[Gould \& Loeb(1992)]{gould92b}
Gould, A., \& Loeb, A.\ 1992, \apj, 396, 104

\bibitem[Gould(2008)]{gould08}
Gould, A.\ 2008, \apj, 681, 1593

\bibitem[Gould et al.(2006)]{gould06}
Gould, A., Udalski, A., An, D., et al. 2006, \apj, 644, L37

\bibitem[Gould et al.(2010)]{gould10}
Gould, A., Dong, Subo, Gaudi, B.~S, et al. 2010, \apj, 720, 1073

\bibitem[Ida \& Lin(2005)]{ida05}
Ida, S., \& Lin, D.~N.~C.\ 2005, \apj, 626, 1045

\bibitem[Kains et al.(2013)]{kains13}
Kains, N., Street, R. A., Choi, J.-Y., et al. 2013, \aap, 552, A70

\bibitem[Kayser et al.(1986)]{kayser86}
Kayser, R., Refsdal, S., \& Stabell, R.\ 1986, \aap, 166, 36

\bibitem[Kervella et al.(2004)]{kervella04a} 
Kervella, P., Th{\'e}venin, F., Di Folco, E., \& S{\'e}gransan, D.\ 2004, \aap, 426, 297 

\bibitem[Kervella \& Fouqu{\'e}(2008)]{kervella-fouque08} 
Kervella, P., \& Fouqu{\'e}, P.\ 2008, \aap, 491, 855

\bibitem[Kubas et al.(2012)]{kubas12}
Kubas, D., Beaulieu, J.-P., Bennett, D.~P., et al.\ 2012, \aap, 540, 78

\bibitem[Laughlin et al.(2004)]{laughlin04}
Laughlin, G., Bodenheimer, P., \& Adams, F.~C.\ 2004, \apj, 612, L73

\bibitem[Mao \& Paczy\'nski(1991)]{mao91}
Mao, S., \& Paczy\'nski, B.\ 1991, \apj, 374, L37

\bibitem[Marcy et al.(1998)]{marcy98}
Marcy, G.\ W., Butler, R.\ P., Vogt, S.\ S., Fischer, D., \& Lissauer, J.~J.\ 1998, \apj, 505, L147

\bibitem[Montet et al.(2014)]{montet14} 
Montet, B.~T., Crepp, J.~R., Johnson, J.~A., Howard, A.~W., \& Marcy, G.~W.\ 2014, \apj, 781, 28 

\bibitem[Nataf et al.(2010)]{nataf10} Nataf, D.~M., Udalski,
A., Gould, A., Fouqu{\'e}, P., \& Stanek, K.~Z.\ 2010, \apjl, 721, L28

\bibitem[Nataf et al.(2013)]{nataf13} Nataf, D.~M., Gould, A., 
Fouqu{\'e}, P., et al.\ 2013, \apj, 769, 88 

\bibitem[Paczy{\'n}ski(1986)]{paczynski86}
Paczy{\'n}ski, B. 1986, \apj, 304, 1

\bibitem[Park et al.(2013)]{park13}
Park, H., Udalski, A., Han, C., et al.\ 2013, \apj, 778, 134 

\bibitem[Pejcha \& Heyrovsk\'y(2009)]{pejcha09}
Pejcha, O., \& Heyrovsk\'y, D. 2009, \apj, 690, 1772

\bibitem[Penny et al.(2011)]{penny11}
Penny, M.~T., Mao, S., \& Kerins, E.\ 2011, \mnras, 412, 607

\bibitem[Poleski et al.(2014)]{poleski14}
Poleski, R., Udalski, A., Dong, S., et al. 2014, \apj, 782, 47

\bibitem[Schechter et al.(1993)]{schechter93}
Schechter, P. L., Mateo, M., \& Saha, A. 1993, \pasp, 105, 1342

\bibitem[Schneider \& Weiss(1986)]{schneider86}
Schneider, P., \& Weiss, A.\ 1986, \aap, 164, 237

\bibitem[Shin et al.(2011)]{shin11}
Shin, I.-G., Udalski, A., Han, C., et al. 2011, \apj, 735, 85

\bibitem[Shvartzvald \& Maoz(2012)]{shvartzvald12} 
Shvartzvald, Y., \& Maoz, D.\ 2012, \mnras, 419, 3631 

\bibitem[Shvartzvald et al.(2014)]{shvartzvald14}
Shvartzvald, Y., Maoz, D., Kaspi, S., et al.  2014, \mnras, 439, 604

\bibitem[Skowron et al.(2011)]{skowron11}
Skowron, J., Udalski, A., Gould, A., et al. 2011, \apj, 738, 87

\bibitem[Skowron \& Gould(2012)]{skowron12} 
Skowron, J., \& Gould, A.\ 2012, arXiv:1203.1034 

\bibitem[Sumi et al.(2010)]{sumi10}
Sumi, T., Bennett, D.~P., Bond, I.~A., et al. 2010, \apj, 710, 1641

\bibitem[Sumi et al.(2011)]{sumi11}
Sumi, T., Kamiya, K., Bennett, D.~P., et al. 2011, Nature, 473, 349

\bibitem[Street et al.(2013)]{street13}
Street, R.~A., Choi, J.-Y., Tsapras, Y., et al. 2013, \apj, 763, 67

\bibitem[Tsapras et al.(2003)]{tsapras03} Tsapras, Y., Horne, K., 
Kane, S., \& Carson, R.\ 2003, \mnras, 343, 1131 

\bibitem[Tsapras et al.(2009)]{tsapras09}
Tsapras, Y., Street, R., Horne, K., et al.\ 2009, Astron. Nachr., 330, 4

\bibitem[Tsapras et al.(2014)]{tsapras14}
Tsapras, Y., Choi, J.-Y., Street, R. A., et al. 2014, \apj, 782, 48

\bibitem[Udalski(2003)]{udalski03}
Udalski, A. 2003, Acta Astron., 53, 291

\bibitem[Udalski et al.(2005)]{udalski05}
Udalski, A., et al. 2005, \apj, 628, L109 

\bibitem[Wambsganss(1997)]{wambsganss97}
Wambsganss, J.\ 1997, \mnras, 284, 172

\bibitem[Yoo et al.(2004)]{yoo04}
Yoo, J., DePoy, D.~L., Gal-Yam, A., et al. 2004, \apj, 603, 139

\end{thebibliography}
\end{document}